# Enhanced thermoelectric figure-of-merit in boron-doped SiGe thin films by nanograin boundaries

Jianbiao Lu,[a] Ruiqiang Guo,[a] Weijing Dai[a] and Baoling Huang[a,b,†]

Boron-doped polycrystalline silicon-germanium (SiGe) thin films are grown by low-pressure chemical vapor deposition (LPCVD) and their thermoelectric properties are characterized from 120 K to 300 K for the potential applications in integrated microscale cooling. The naturally formed grain boundaries are found to play a crucial role in determining both the charge and thermal transport properties of the films. Particularly, the grain boundaries create energy barriers for charge transport which lead to abnormal dependences of charge mobility on doping concentration and temperature. Meanwhile, the unique columnar grain structures result in remarkable thermal conductivity anisotropy with the in-plane thermal conductivities of SiGe films about 50% lower than the cross-plane values. By optimizing the growth conditions and doping level, a high figure of merit ($ZT$) of 0.2 for SiGe films is achieved at 300 K, which is about 100% higher than the previous record for p-type SiGe alloys, mainly due to the significant reduction in the in-plane thermal conductivity caused by nanograin boundaries. The low cost and excellent scalability of LPCVD render these high-performance SiGe films ideal candidates for thin-film thermoelectric applications.

[a] *Department of Mechanical and Aerospace Engineering, The Hong Kong University of Science and Technology, Clear Water Bay, Kowloon, Hong Kong.*
[b] *The Hong Kong University of Science and Technology Shenzhen Research Institute, Shenzhen, China*
† *Author to whom correspondence should be addressed. Electronic mail: mebhuang@ust.hk.*

**Introduction**

Efficient microscale cooling is becoming highly attractive for microelectronic and optoelectronic applications due to its capability of stabilizing the device temperature, improving the reliability, and reducing noise levels[1]. As one of the few most promising techniques that can be downscaled to micro domain, thin-film thermoelectric (TE) refrigeration provides an all-solid-state cooling solution, which offers a compact structure, fast thermal response, high reliability and long lifetime[2]. The efficiency of thermoelectric devices relies on the performance of TE materials, which is often evaluated by the dimensionless figure-of-merit $ZT = S^2\sigma T/(k_L + k_e)$, where $S$ denotes the Seebeck coefficient, $T$ is the temperature, $\sigma$ is the electrical conductivity, and $k_e$ and $k_L$ are the electrical and lattice thermal conductivities, respectively. Despite their high bulk $ZT$ value (~1.0 at 300 K), conventional Te-based thermoelectric thin films, such as $Bi_2Te_3$ films, are not quite appropriate for the integrated electronic cooling due to their high toxicity, low mechanical strength, and incompatibility with standard IC fabrication processes[2]. Meanwhile, the multiple deposition processes and poor electrical contacts will severely deteriorate the overall performance of devices consisting of these Te-based films and may result in an effective $ZT$ below 0.07[2]. As one of the most widely used Si-based high-temperature thermoelectric materials, SiGe can overcome these problems and become a promising candidate for micro TE devices, although its room-temperature TE performance is relatively poor. While the room-temperature $ZT$ value of bulk SiGe is typically below 0.1[3], even lower $ZT$ values (0.02~0.06) or incomprehensive TE properties are often reported for SiGe films fabricated using various techniques ranging from atmospheric pressure chemical vapor deposition (APCVD)[4], plasma enhanced chemical vapor deposition (PECVD)[5], low-pressure chemical vapor deposition (LPCVD)[6], to sputtering[7] and molecular beam deposition[8]. Further efforts to improve the $ZT$ values of SiGe thin films continue.

In recent years, significantly improved $ZT$ values have been reported for nanostructured SiGe such as SiGe nanocomposites[9, 10] and nanowires[11], mainly due to the reduction of lattice thermal conductivity by boundary scattering. These nanocomposites and nanowires, however, are difficult to apply in thin film applications because of the challenges in uniformity control and integration. It is expected that the grain boundaries may also help to improve TE performance of SiGe thin films. In this work, we have investigated the thermoelectric properties of boron-doped SiGe thin films grown by LPCVD, a cost-effective technique offering good quality control on large-area deposition and excellent scalability, under different growth conditions and doping treatments. It is found that the naturally formed columnar grains in polycrystalline SiGe films grown by LPCVD do play an important role in determining the thermal and charge transport in the films. Through optimizing the growth condition and doping concentration, a high in-plane *ZT* value of 0.2 for p-type SiGe films at 300 K is achieved, which is even more than 100% higher than the previous record for bulk SiGe. The *ZT* improvement is mainly attributed to the reduction of in-plane thermal conductivity by the grain boundaries. However, the columnar grains seem to have minor effects on the cross-plane thermal transport.

**Sample preparation and characterization**

The SiGe thin films were grown by LPCVD on a <100>-oriented silicon wafer coated with 1 μm thermally-grown wet $SiO_2$. A 3-nm amorphous Si buffer layer was first deposited on the substrate to prevent the island growth of the SiGe layer. Then 0.7~1 μm SiGe films were grown at 600 mTorr and preselected deposition temperatures (550 ~ 650 °C) using $SiH_4$ and $GeH_4$ gases with flow rates ranging from 100 SCCM and 7 SCCM to 150 SCCM and 6 SCCM, respectively to achieve a Ge concentration of about 30%. After being capped by low stress nitride (LSN), the SiGe films were properly doped with boron using ion implantation techniques. The samples were then annealed to recrystallize the SiGe films and activate the dopants. Different deposition temperatures, doping doses and annealing treatments were tested during the sample preparation in order to optimize the thermoelectric properties of the SiGe films. The samples were characterized with X-ray photoelectron spectroscopy (XPS) to confirm the Si to Ge ratio to be around 74:26, secondary ion mass spectrometry (SIMS) to verify the homogeneous composition distribution. The surface morphology and cross-sectional structure of the annealed thin films were analyzed by scanning electron microscopy (SEM) and transmission electron microscopy (TEM) (Fig. 1(a)), showing a columnar grain structure with a grain size slightly increasing from the bottom to the top, like bundles of nanorods packed together, similar to the work reported by Takashiri[6] and McConnell[12]. Different deposition temperatures as well as the post annealing may affect the polycrystalline structure formation. Figure 1(b) shows the high-resolution TEM images of the grains in films grown at 650 ℃. These grains are single crystalline but with different lattice orientations, illuminating the polycrystalline nature of the films. The inset shows the diffraction pattern of a single grain, further confirming its good crystallinity. Figure 1(c) shows the X-ray diffraction (XRD) patterns of SiGe films grown at 550 ℃ and 650 ℃ before and after 60 min's annealing at 1000 ℃ in nitrogen environment. Before the annealing, the sample grown at 650 ℃ shows much better crystallinity than the one grown at 550 ℃. However, the crystallinity difference becomes much smaller after the annealing. The well-crystallized samples show a (111)-preferred orientation, similar to the high-performance nanostructured bulk SiGe prepared by hot pressing[9] except for the better crystallinity. The grain size distributions of the samples grown at 650 ℃ and then annealed with different durations were studied by the scanning electron microscope (SEM), as shown in Fig. 1(d). It is clear that with the increase of the annealing time, the grains became more uniform and the average grain size increased from 231 nm to 291 nm.

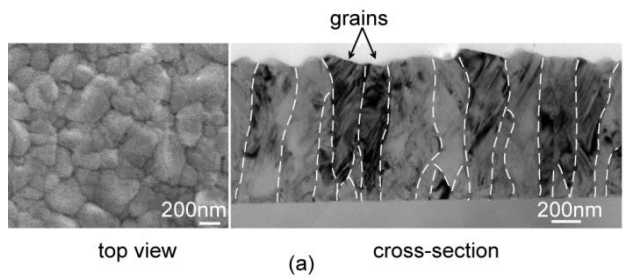

(a) top view / cross-section

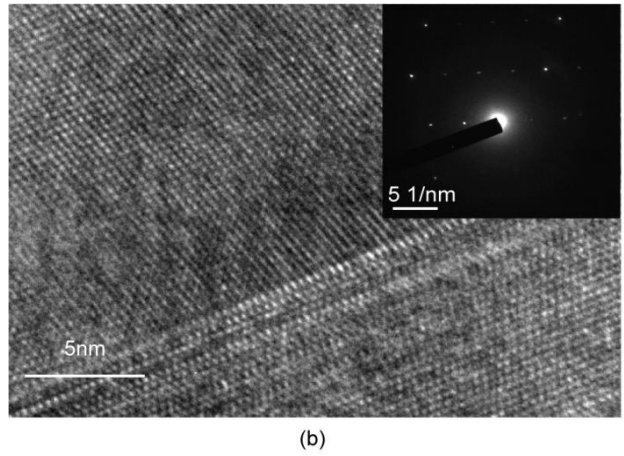

(b)

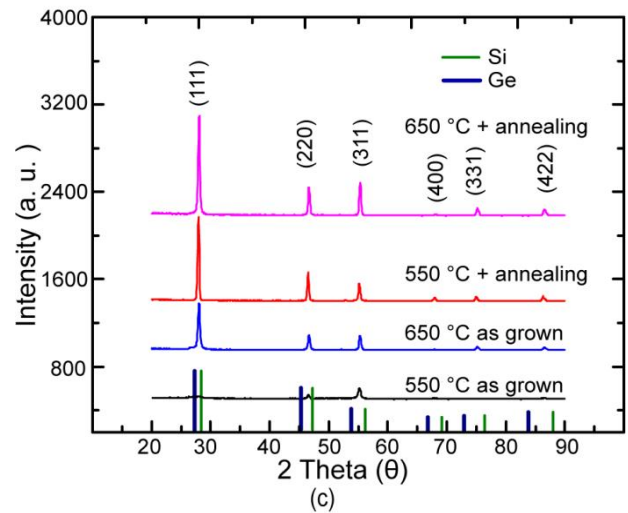

(c)

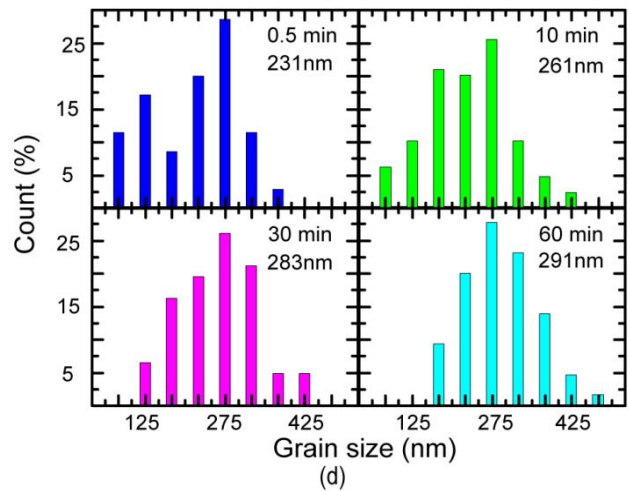

(d)

Fig. 1. (a) SEM image of the top surface morphology and TEM image of cross-sectional structure of a polycrystalline SiGe thin film grown by LPCVD at 650 ℃. Some grain boundaries are sketched using dashed lines. (b) High-resolution TEM image of grains and grain boundaries. The inset picture shows the diffraction pattern of a grain. (c) X-ray diffraction patterns of SiGe thin films grown at 550 and 650 ℃ before and after 60 min's annealing at 1000 ℃ in nitrogen. The vertical bars on the *x* axis indicate the corresponding patterns of Si and Ge single crystals. (c) The evolution of the grain size distribution of the SiGe thin film with the annealing time, varying from 0.5 to 60 min.

*Measurements and results*

To avoid the error introduced by sample variance, all the thermoelectric properties, including $S$, $\sigma$ and $k$, were measured on the same sample. 5-8 samples in each batch were used for the characterization and the measurement uncertainties were estimated from the deviation of the sample properties from the average values. During the measurements, the samples were placed in a vacuum chamber (Lakeshore TTPX probe station) with a pressure below $10^{-5}$ Torr to avoid possible heat loss to the ambient. The measurement techniques used are schematically illustrated in Fig. 2. The electrical conductivity ($\sigma$) were measured using a Van der Pauw configuration[13]. The same configuration was used to measure the Hall mobility of the charges by applying a magnetic field. To get the Seebeck coefficient, one end of the sample was suspended and heated by a platinum (Pt) coil heater, while the other end was fixed on a copper heat sink. The temperature difference $\Delta T$ between two locations on the film surface was recorded by two Type T thermocouples; meanwhile, the corresponding voltage $U$ between the two points was measured by a nano voltmeter (Keithley 2182A). By changing the input heating power, a series of $\Delta T$ and $U$ were obtained and the Seebeck coefficient of the film was then fitted according to $S-S_m = -U/\Delta T$, where $S_m$ is the Seebeck coefficient of the metal (Cu) electrodes. The thermal conductivities of SiGe films were measured using the two-wire differential $3\omega$ method[14-18]. Two Pt line heaters with linewidths of 2 μm and 20 μm respectively were deposited on the sample but electrically insulated from the film by a 50-nm $Al_2O_3$ layer deposited with atomic layer deposition (ALD). A reference sample of the same structure but without the SiGe film was fabricated simultaneously under the same conditions and it was used as a reference to subtract the unknown properties of the underneath multilayer structure. The cross-plane thermal conductivities were first extracted using the differential method according to the signals from the wide heater. The in-plane thermal conductivities ($k_{in}$) of the thin films were then extracted from the temperature response of the two line heaters under input currents of different frequencies according to the general 2-D model for heat transfer in multilayer structures[15, 18].

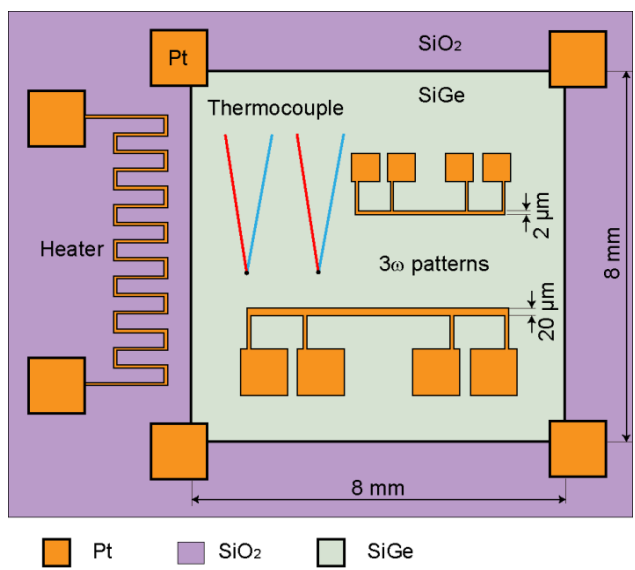

Fig. 2. Schematics for the characterization of the thermoelectric properties of SiGe thin films.

*Effects of annealing time on thermoelectric properties*

We first studied the effects of annealing treatment on the thermoelectric properties of SiGe thin films. As already shown in Fig. 1(d), a longer annealing period leads to a larger average grain size, which will reduce grain boundary density and might affect the thermoelectric transport properties of interest. A batch of samples deposited at 650 ℃ and then implanted with boron at a fixed level of $8 \times 10^{15}$/cm$^2$ were annealed at ~1000 ℃ in nitrogen atmosphere for 30 sec, 10 min, 30 min and 60 min, respectively. The measured thermoelectric properties of these samples at 300 K are shown in Fig. 3.

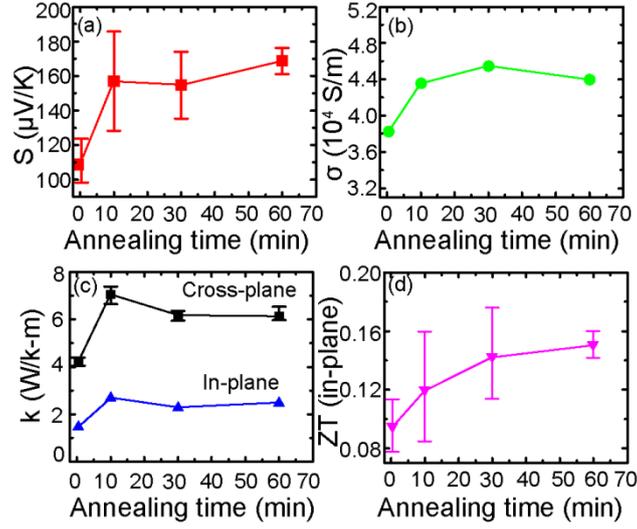

FIG. 3. Effects of the annealing time on the (a) Seebeck coefficient, (b) thermal conductivity, (c) electrical conductivity, and (d) figure of merit of LPCVD-grown SiGe films doped with boron at a fix level of $8 \times 10^{15}$ /cm$^2$ at 300 K.

The annealing treatment in the current work does not result in appreciable changes of the free carrier concentration in the samples, as confirmed in the Hall effect measurements. With the increase of annealing time, $S$ first increases but changes little after being annealed for 10 min or longer, probably due to improved crystallinity after long-time annealing. Also, it seems that increasing the annealing time may reduce the sample-to-sample variation. Figure 3(b) shows that $\sigma$ takes a similar trend. When the annealing time increases from 30 sec to 60 min, $\sigma$ of the samples are improved by ~ 15%, mainly due to the enhanced mobility benefitting from the improvement in crystallinity and the reduction of the grain boundary density. The thermal conductivity $k$ also increases slightly with the annealing time due to less boundary scatterings for phonons and the increase of electrical thermal conductivity. Both the cross-plane values $k_{cr}$ and in-plane ones $k_{in}$ are much lower than the typical values of large-grain bulk SiGe with similar compositions (~8-10 W/m-K)[19]. Considering the fact that the dominant phonon mean free path (MFP) in bulk SiGe is more than 300 nm[20, 21], the small grain sizes of the samples (~100-400 nm) or even the confined thicknesses (~ 1μm) can efficiently suppress the phonon transport in SiGe thin films. Interestingly, the cross-plane thermal conductivities $k_{cr}$ are consistently more than 100% higher than the in-plane values $k_{in}$, as opposite to the phenomena in single-crystalline thin films, where $k_{cr}$ is often smaller than if not equal to $k_{in}$. This is probably attributed to the unique columnar structure of the thin films, since the phonon scattering is more severe in the in-plane direction due to the presence of perpendicular grain boundaries. Similar phenomena have been reported for CVD-grown diamond films[22]. Although with a relatively large uncertainty, the calculated sample-averaged ZT values in the in-plane direction (Fig. 3(d)) seems to increase with the increasing annealing time, leading to the conclusion that a bigger grain size in this grain size range will benefit ZT mainly due to the enhancement of the power factor $S^2\sigma$.

*Effects of doping concentration on thermoelectric properties*

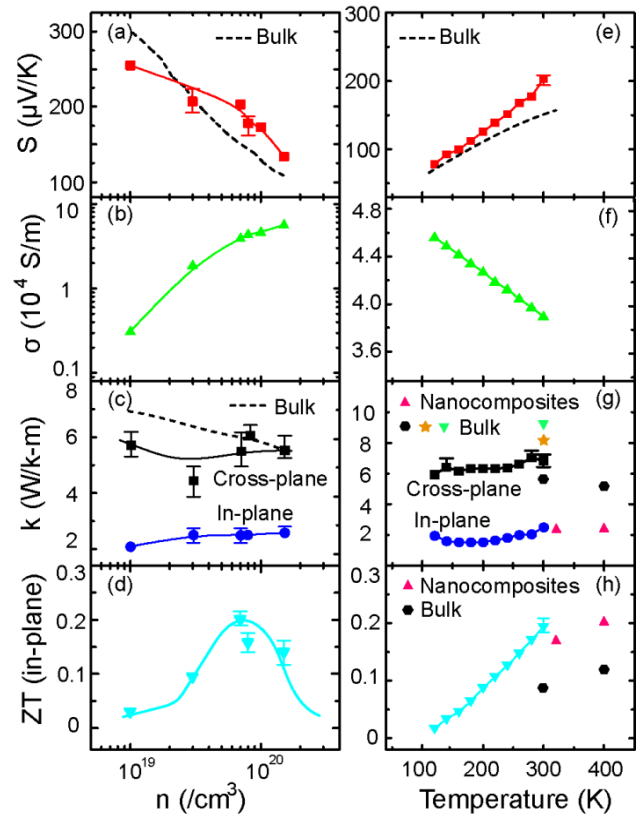

FIG. 4. Effects of the doping concentration on the (a) Seebeck coefficient (b) electrical conductivity (c) thermal conductivity and (d) figure of merit of SiGe samples at 300K. The temperature dependences of the (e) Seebeck coefficient, (f) electrical conductivity, (g) thermal conductivity, and (h) figure of merit of the SiGe sample with a doping concentration of $7\times10^{19}$ /cm$^3$ are shown together with those of bulk SiGe[19, 23-25] and SiGe nanocomposites[9] reported in the literature. The dashed lines and separate points are taken from references.

Keeping the optimal growth condition and thermal treatment, we then explored the effects of doping concentration on the thermoelectric properties of SiGe thin films to optimize their performance. A batch of samples grown at 650 ℃ were implanted with different boron concentrations $n$, and then all were annealed in nitrogen at 1000 ℃ for 60 minutes. The thermoelectric properties of SiGe films at 300 K were then plotted with respect to $n$ in Figs. 4(a)-(d). Similar to other thermoelectric materials, the Seebeck coefficients of SiGe films decrease with the increase of $n$ but the dependence seems relatively weaker than the bulk counterpart[26] (dashed line). When the doping concentration is relatively smaller, $S$ of the films is ~15% smaller than the bulk value, which is probably due to the better crystallinity in the bulk sample. However, when the doping concentration is higher than $5 \times 10^{19}$ /cm$^3$, $S$ of the films is 10% ~ 30% higher than the bulk value. This is probably due to the aggregation of dopants on the grain boundaries in our samples, which leads to a lower dopant activation ratio[27]. In the Hall effect measurement, it is confirmed that at $n \sim 1 \times 10^{19}$/cm$^3$, the dopants are almost totally activated, while at $n \sim 1.5 \times 10^{20}$/cm$^3$ the activated dopants account for only 65% of the total doping amount. Consequently, the variation of $\sigma$ with $n$ also becomes gentler at a higher doping concentration, although the mobility still increases slightly at the same time. Once again, the thermal conductivities of SiGe thin films show a strong anisotropy, e.g., the cross-plane thermal conductivities are generally 1~2 times higher than the in-plane values. $k_{cr}$ does not change significantly with the doping concentration because of the dominant alloy scattering and the fact that the dopants will preferentially reside at the grain boundaries[9, 27] and will not create remarkable defects inside the grains. On the other hand, $k_{in}$ increases about 30% when the doping concentration increases from $1 \times 10^{19}$/cm$^3$ to $1.5 \times 10^{20}$/cm$^3$, mainly due to the increase of electrical thermal conductivity according to the Wiedemann-Franz law[28]. The overall effect of the doping on the thermoelectric performance $ZT$ is then evaluated. Figure 4(d) illustrates that the optimum doping concentration is in the vicinity of $7 \times 10^{19}$ /cm$^3$, with the in-plane $ZT$ value approaching 0.2 at 300 K, which is more than 100% higher than the state-of-the-art values of 0.05~0.1 for bulk p-type SiGe[23, 29] and comparable with the previous record (about 0.18 at 300 K) for the high-performance SiGe nanocomposites[9].

For many cooling applications with SiGe thin films, their low temperature behaviors are important. Figures 4(e)-(h) show the measured temperature dependences of thermoelectric properties of the optimized samples with an optimum doping concentration of $7 \times 10^{19}$ /cm$^3$. Compared to the bulk samples[26] with a similar composition but a slightly higher doping concentration, below 200 K the $S$ values are very close, but $S$ of the SiGe films increases faster than the bulk counterpart and similar to the results reported for LPCVD-grown $Si_{0.8}Ge_{0.2}$ film[6]. This might be also due to the trapping of dopants at grain boundaries in the films, which leads to a lower hole concentration at the same temperature. On the other hand, $\sigma$ of the SiGe sample decreases with the increasing temperature mainly due to the decrease of the mobility $\mu$ caused by stronger charge-phonon scattering. For the thermal transport, both $k_{cr}$ and $k_{in}$ increase gently with elevated temperatures. By applying the Wiedemann-Franz law[28] and using the revised Lorenz number for Si[30], the electrical contribution to $k_{in}$ was estimated to be ~ 10% at 300 K. The $ZT$ of the optimized films shows a monotonically increasing trend with the increase of temperature and reaches 0.2 at 300 K, remarkably outperforms their bulk counterparts [9, 23, 26, 29]. It is worth noting that $ZT$ of the thin films reaches 0.1 at around 230 K. Considering the fact that the $ZT$ values of bulk SiGe generally decrease at lower temperatures, i.e., their $ZT$ values would be even lower than the room temperature value (< 0.1) below 300 K, SiGe thin films can deliver a superior performance over bulk SiGe for refrigeration cooling.

**Summary**

We have successfully synthesized high-performance SiGe thin films through LPCVD by optimizing the growth conditions and doping concentration. The synthesized SiGe films are polycrystalline and have a columnar grain structure. Important thermoelectric properties of the SiGe films, including electrical conductivity, Seebeck coefficient, mobility and thermal conductivity, have been characterized from 120 to 300 K. It is found that the energy barriers at the grain boundaries are found to play a crucial role in determining the charge transport properties while the columnar grain structure can substantially reduce the in-plane thermal conductivity but has a negligible influence on the cross-plane transport. A high *ZT* value of 0.2 has been achieved for optimized samples and 300 K and the improvement is mainly due to the thermal conductivity reduction caused by the naturally formed columnar grain structures in the films. Considering the high throuhout, scalability and low cost of LPCVD technique, LPCVD-grown high-*ZT* SiGe thin films may be ideal candidates for the thin-film thermoelectric applications.

## Acknowledgement

The authors are thankful for the support from the Hong Kong General Research Fund (Grant Nos. 613211 and 623212) and National Natural Science Foundation of China (Grant No. 51376154).